\documentclass{article}
\usepackage{hyperref}
\title{Informal Resource Letter -- Nonlinear quantum mechanics on arXiv up to August 2004}
\author{George Svetlichny\footnote{Departamento de Matem\'atica, Pontif\'{\i}cia Universidade Cat\'olica, Rio de Janeiro, Brazil \newline
svetlich@mat.puc-rio.br \hfill \url{http://www.mat.puc-rio.br/\~svetlich}}}
\begin{document}
\maketitle
\begin{abstract}
I compiled a list of articles on arXiv that deal with possible fundamental quantum nonlinearities or examine the origins of its linearity. The list extends til August 2004.
\end{abstract}
\section{Introduction} As a result of a search for articles dealing with nonlinear quantum theories I compiled a list of those available on arXiv from the inception of the archives until August 2004. Thinking that such a list may be of use to others, I decided to post it on the archives. In this compilation I only included articles that deal with supposed fundamental nonlinearities in the quantum formalism itself and left out those that deal with nonlinear problems within normal linear quantum contexts and those that deal with nonlinear Schr\"odinger equations (or similar) as phenomenological theories which do not question the fundamental linearity of quantum physics. Included also are articles that  examine the origins of quantum linearity. As this was an informal search I make no claim to completeness nor rigor in my selection criteria. I apologize for any references that were not included and for those included wrongly. Likewise no attempt was made to classify the articles as to topic or viewpoint. The list is in chronological order according to the arXiv identification number. Jornal references are given to the extend supplied by the authors on arXiv.

Nonlinear quantum mechanics is an idea that has willingly or unwillingly (I'm in the unwilling category) intrigued many researchers. In a sense, its time has come. With a history of thirty odd years, it has provided us with important insights into the structure of our world and of the theories we build of them, and may still provide us with viable alternatives as we try to explore the realms of quantum geometry, cosmology, and  ultra-high energy. It should engender new mathematical ideas which could provide interesting models for other sciences such as, biology,  psychology, economics, etc.  A true resource letter including the pre-internet literature and that not found on arXiv would be most helpful in making  accessible  all the fascinating ideas that have come out of this research, and in giving proper recognition to the people that have contributed to it. Hopefully this informal letter will be a stimulus to the creation of such a list.

This list will be updated in so far as omissions or inaccuracies are pointed out by readers. No material posted after August 2004 will be included.

\section*{Acknowledgements}
My thanks to professors Gian Paolo Beretta, Marek Czachor, Mauro Gatti, Tamas Geszti, Jorge Mahecha, Tim Palmer, and Arun Pati for their contributions toward improving this list. This work received partial financial support from the Conselho Nacional de Desenvolvimento Cient\'{\i}fico e Tecnol\'ogico (CNPq), and the Funda\c{c}\~ao Carlos Chagas Filho de Amparo \`a Pesquisa do Estado do Rio de Janeiro  (FAPERJ).


\begin{thebibliography}{xxx}

\bibitem{czac9406174}
Marek~Czachor, ``Elements of nonlinear quantum mechanics (Part II): Triple bracket generalization of quantum mechanics", hep-th/9406174.

\bibitem{cass9409003} M.~J.~Cassidy, ``Nonlinearity in Quantum Theory and Closed Timelike Curves", {\sl Phys.~Rev.}\ {\bf D52} (1995) 5676, (gr-qc/9409003).

\bibitem{czac9501007}Marek~Czachor, ``Nonlinear Schroedinger equation and two-level atoms", {\em Phys.~Rev.\/} {\bf A53} 1310 (1996), (quant-ph/9501007).

\bibitem{czac9501008}Marek~Czachor,
``Triple bracket generalization of quantum mechanics", quant-ph/9501008.


\bibitem{do-go-na9502014}H.-D.~Doebner, G.~A.~Goldin, and P.~Nattermann,
``A FAMILY OF NONLINEAR SCHR\"ODINGER EQUATIONS: LINEARIZING TRANSFORMATIONS AND RESULTING  STRUCTURE",
 in J.-P.~Antoine et.al. (Eds.), {\sl Quantization, Coherent States, and Complex Structures \/}, p. 27, Plenum 1995 (ISBN 0-306-45214-6), (quant-ph/9502014).

\bibitem{luec9505022}W.~Luecke, ``Nonlinear Schroedinger Dynamics and Nonlinear Observables", quant-ph/9505022.

\bibitem{palm9505025}T.~N.~Palmer, ``A Local Deterministic Model of Quantum Spin Measurement", quant-ph/9505025.

\bibitem{na-sc9506033}P.~Nattermann and  W.~Scherer,
``Nonlinear Gauge Transformations and Exact Solutions of the Doebner-Goldin Equation",
in  H.-D.~Doebner et.al. (Eds.),
{\sl Nonlinear, Deformed, and Irreversible Quantum Systems}, p. 188, World Scientific 1995 (ISBN 981-02-2266-1), (quant-ph/9506033).

\bibitem{jone9507001}K. R. W. Jones, ``Newtonian Quantum Gravity", {\sl Austral.~J.~Phys.} {\bf 48} (1995) 1055, (quant-ph/9507001).

\bibitem{na-zh9510001}P.~Nattermann and R.~Zhdanov,
``On Integrable Doebner-Goldin Equations",
{\sl J.~Phys.} {\bf A29} (1996) 2869, (solv-int/9510001).

\bibitem{svet9511002}George~Svetlichny,
``Quantum Formalism with State-Collapse and Superluminal Communication",
{\sl Foundation of Physics} {\bf 28} (1998) 131,
(quant-ph/9511002).

\bibitem{svet9512004} George~Svetlichny,
``Quantum Evoluton and Space-time Structure",
quant-ph/9512004.

\bibitem{czac9601015}Marek~Czachor,
``Nambu-Type Generalization of the Dirac Equation",
{\sl Phys.~Lett.} {\bf A225} (1997) 1, (quant-ph/9601015).

\bibitem{sc-us9602012}A.~Scotti and A.~Ushveridze, ``Non-linear Quantization of Integrable Classical Systems", {\sl J.~Math.~Phys.} {\bf 38} (1997) 4073,
(quant-ph/9602012).


\bibitem{meye9605023}David~A.~Meyer, ``Unitarity in one dimensional nonlinear quantum cellular automata", quant-ph/9605023.

\bibitem{leif9610030}Peter~Leifer,
``Superrelativity as a unification of quantum theory and relativity",
quant-ph/9610030.

\bibitem{leif9610043}Peter~Leifer,
``Quantum theory requires gravity and superrelativity",
gr-qc/9610043.


\bibitem{cz-ku9612031}M.~Czachor and M.~Kuna, ``Off-shell indefinite-metric triple-bracket generalization of the Dirac equation ", in {\sl Group 21 - Physical Applications and Mathematical Aspects of Geometry,
Groups, and Algebras\/}, edited by H.-D.~Doebner, P.~Nattermann, and W.~Scherer, p.
451 (World Scientific, Singapore, 1997), (quant-ph/9612031).


\bibitem{leif9702160}Peter~Leifer, ``Nonlinear modification of quantum mechanics", hep-th/9702160.


\bibitem{natt9703017}Peter~Nattermann, ``Generalized Quantum Mechanics and Nonlinear Gauge Transformations", in Symmetry in Science, Vol. IX, edited by B.~Gruber and M.~Ramek (Plenum, New York, 1997), p. 269,  (quant-ph/9703017).

\bibitem{ah-re9704001}Y.~Aharonov and B.~Reznik,
``Measurability in Linear and Non-Linear Quantum Mechanical Systems",
quant-ph/9704001.

\bibitem{do-ko9705227}  G.~Domokos, and S.~Kovesi-Domokos,  ``Tests of Basic Quantum Mechanics in Oscillation Experiments", {\sl J.Phys.\/} {\bf A32} 4105 (1999), (hep-ph/9705227).

\bibitem{as-sc9706069}Abhay~Ashtekar, Troy~A.~Schilling, ``Geometrical Formulation of Quantum Mechanics", in {\sl On Einstein's Path\/}, A. Harvey (Ed.), Springer, Berlin, 1998, (gr-qc/9706069).

\bibitem{cz-ma9707013}M.~Czachor and M.~Marciniak, ``Density matrix interpretation of solutions of Lie-Nambu equations", {\sl Phys.~Lett.} {\bf A239} 353 (1998), (quant-ph/9707013).

\bibitem{lu-na9707055}W.~Luecke and P.~Nattermann,
``Nonlinear Quantum Mechanics and Locality", in Symmetry in Science, Vol. X, edited by B.~Gruber and M.~Ramek (Plenum, New York, 1998), p. 197,
(quant-ph/9707055).

\bibitem{cz-ku9708029}M.~Czachor and M.~Kuna, ``Complete positivity of nonlinear evolution: A case study",
{\sl Phys.~Rev.\/} {\bf A58} 128 (1998), (quant-ph/9708029).

\bibitem{be-hu-gi9708040}H.~Bechmann-Pasquinucci, B.~Huttner, and N.~Gisin,
``Nonlinear quantum state transformation of spin-1/2, {\sl Phys.~Lett.} {\bf A242} (1998) 198, (quant-ph/9708040).


\bibitem{czac9708052}Marek~Czachor,
``Nonlocal looking equations can make nonlinear quantum dynamics local",
{\sl Phys.~Rev.} {\bf A57} (1998) 4122,
(quant-ph/9708052).

\bibitem{do-go-na9709036}H.-D.~Doebner, G.~A.~Goldin, P.~Nattermann, ``Gauge Transformations in Quantum Mechanics and the Unification of Nonlinear Schr\"odinger Equations", {\sl J.~Math.~Phys.} {\bf 40} (1999) 49, (quant-ph/9709036).

\bibitem{pusz971007}Waldemar~Puszkarz,
``Higher Order Modification of the Schroedinger Equation",
quant-ph/9710007.

\bibitem{pusz971008}Waldemar~Puszkarz,
``Relativistically Extended Modification of the Schroedinger Equation",
quant-ph/9710008.

\bibitem{pusz971009}Waldemar~Puszkarz,
``Extension of the Staruszkiewicz Modification of the Schroedinger Equation",
quant-ph/9710009.

\bibitem{pusz9710010}Waldemar~Puszkarz,
``Nonlinear Phase Modification of the Schroedinger Equation",
quant-ph/9710010.

\bibitem{luec9710033}W.~Luecke,
``Gisin Nonlocality of the Doebner-Goldin 2-Particle Equation",
 quant-ph/9710033.

\bibitem{maha9710040}Karmadeva~Maharana,
``Nonlinear Dirac and diffusion equations in $1 + 1$ dimensions from stochastic considerations",
{\sl Phys.~Rev.} {\bf E62} (2000) 1683, (physics/9710040).

\bibitem{czac9711053}Marek~Czachor,
``Structure of nonlinear gauge transformations",
{\sl Phys.~Rev.} {\bf A57} (1998) 2263, (quant-ph/9711053).

\bibitem{czac9711054}Marek~Czachor,
``Lie-Nambu and beyond",
{\sl Int.~J.~Theor.~Phys.} {\bf 38} (1999) 475, (quant-ph/9711054).

\bibitem{leif9711059}Peter~Leifer, ``Dynamical Spacetime and the
Curvature of Projective State Space", gr-qc/9711059.

\bibitem{hubs9712047}Tristan~H\"ubsch,
``Quantum Mechanics is Either Non-Linear Or Non-Introspective",
{\sl Mod.~Phys.~Lett.} {\bf A13} (1998) 2503, (quant-ph/9712047).



\bibitem{ab-ll9801041}Daniel~S.~Abrams and Seth~Lloyd,
``Nonlinear quantum mechanics implies polynomial-time solution for NP-complete and \#P problems",
{\sl Phys.~Rev.~Lett.} {\bf 81} (1998) 3992, (quant-ph/9801041).


\bibitem{pusz9802001}Waldemar~Puszkarz,
``Energy Ambiguity in Nonlinear Quantum Mechanics", quant-ph/9802001.


\bibitem{czac9802051}Marek~Czachor,
``Notes on nonlinear quantum algorithms",
{\sl Acta~Phys.~Slov.} {\bf 48} (1998) 157, (quant-ph/9802051).


\bibitem{doeb9803011}H.-D.~Doebner,
``Remarks on a Nonlinear Quantum Theory", quant-ph/9803011.

\bibitem{czac9803019}Marek~Czachor,
``Local modification of the Abrams-Lloyd nonlinear algorithm",
quant-ph/9803019.

\bibitem{pati9803082}A.~K.~Pati, ``Geometry of the Hilbert space and the Quantum Zeno Effect",
{\sl Phys.~Rev.\/} {\bf A58} 831 (1998), (quant-ph/9803082).

\bibitem{le-cz9804052}S.~B.~Leble and M.~Czachor, ``Darboux-integrable nonlinear Liouville-von Neumann
equation", {\sl Phys.~Rev.\/} {\bf E58} 7091 (1998), (quant-ph/9804052).

\bibitem{fink9809017}J.~Finkelstein, ``Comment on `Consistency, amplitudes, and probabilities in quantum theory'",
{\sl Phys.~Rev.} {\bf A60} 1723 (1999), (quant-ph/9809017).

\bibitem{cz-na9809061}M.~Czachor and J.~Naudts, ``Microscopic foundation of nonextensive statistics",
{\sl Phys.~Rev.\/} {\bf E59} R2497 (1999), (quant-ph/9809061).

\bibitem{ku-cz-le9810023}M.~Kuna, M.~Czachor, and S.~B.~Leble, ``Nonlinear von Neumann-type equations:
Darboux invariance and spectra", {\sl Phys.~Lett.\/} {\bf A225} 42 (1999), (quant-ph/9810023).

\bibitem{ca-do-mi9811016}E.~C.~Caparelli, V.~V.~Dodonov, and S.~S.~Mizrahi,
``Finite-Length Soliton Solutions of the Local Homogeneous Nonlinear Schroedinger Equation",
{\sl Phys.~Scripta} {\bf 58} 417 (1998), (quant-ph/9811016).

\bibitem{tern9811036}Daniel~R.~Terno, ``Non-linear operations in quantum information theory", {\sl Phys.~Rev.\/} {\bf  A59} 3320 (1999), (quant-ph/9811036).

\bibitem{pusz9903010}Waldemar~Puszkarz,
``On Solutions to the Nonlinear Phase Modification of the Schroedinger Equation", quant-ph/9903010.


\bibitem{luec9904016}W.~Luecke,
``Nonlocality in Nonlinear Quantum Mechanics", in  {\sl Trends in Quantum Mechanics\/}, ed. H.-D. Doebner et al. (eds.), p. 235 (World Scientific, Singapore, 2000),
(quant-ph/9904016).

\bibitem{czac9904110}Marek~Czachor, Maciej~Kuna, Sergiej~B.~Leble, and
Jan~Naudts,
``Nonlinear von Neumann-type equations", in  {\sl Trends in Quantum Mechanics\/}, ed. H.-D. Doebner et al. (eds.), p. 209 (World Scientific, Singapore, 2000), (quant-ph/9904110).


\bibitem{pusz9905046}Waldemar~Puszkarz,
``Non-separability without Non-separability in Nonlinear Quantum Mechanics", quant-ph/9905046.


\bibitem{svet9906117} George~Svetlichny ``Non-linear Schroedinger Equations, Separation and Symmetry",  {\sl J.~Nonlin.~Math.~Phys.} {\bf 2} 2 (1995), (quant-ph/9906117).

\bibitem{cz-do-sy-wa9907059}M.~Czachor, H.-D.~Doebner, M.~Syty, K.~Wasylka, ``Von Neumann equations with time-dependent Hamiltonians and supersymetric quantum meachanics", {\sl Phys.~Rev.\/} {\bf  E61} 3325 (2000), (quant-ph/9907059).

\bibitem{bona9909022}Pavel~Bona,``Extended Quantum Mechanics", {\sl Acta~Phys.~Slov.}
 {\bf 50} (2000) 1, (math-ph/9909022).


\bibitem{bona9910011}Pavel~Bona,
``Geometric Formulation of Nonlinear Quantum Mechanics for Density Matrices",
quant-ph/9910011.



\bibitem{bona9910012}Pavel~Bona,
``On Symmetries in Nonlinear Quantum Mechanics",
quant-ph/9910012.

\bibitem{kubo9911002}Hiroto~Kubotani,
``Quantum Trajectory in Multi-Dimensional Non-Linear System",
quant-ph/9911002.

\bibitem{pusz9912006}Waldemar~Puszkarz,
``On the Staruszkiewicz Modification of the Schroedinger Equation",
quant-ph/9912006.



\bibitem{svet9912099} George~Svetlichny, ``The Space-time Origin of Quantum Mechanics: Covering Law",  {\sl Found.~Phys.} {\bf 30} 1819 (2000), (quant-ph/9912099).




\bibitem{jin0001029}Wei~Min~Jin,
``From time inversion to nonlinear QED",
{\sl Found.~Phys.} {\bf 30} (2000) 1943, (quant-ph/0001029).


\bibitem{gold0002013}Gerald~A.~Goldin,
``Perspectives on Nonlinearity in Quantum Theory",
quant-ph/0002013.


\bibitem{hans0003083}Johan~Hansson,
``Nonlinear gauge interactions - A solution to the `measurement problem' in quantum  mechanics?",
quant-ph/0003083.

\bibitem{lloy0003151}Seth Lloyd,
``Unconventional Quantum Computing Devices",
quant-ph/0003151.


\bibitem{czac0005030}N.~V.~Ustinov, S.~B.~Leble,  M.~Czachor, and M.~Kuna,
``Darboux-integration of \(id\rho/dt=[H,f(\rho)]\)",
{\sl Phys.~Lett.} {\bf A279} (2001) 333, (quant-ph/0005030).

\bibitem{el-ma-na0007044}John~Ellis, N.~E.~Mavromatos, D.~V.~Nanopoulos, ``How Large are Dissipative Effects in Non-Critical Liouville String Theory?", {\sl Phys.~Rev.\/} {\bf D63} 024024 (2001), (gr-qc/0007044).


\bibitem{gheo0007111}S.~Gheorghiu-Svirschevski,
``Nonlinear quantum evolution with maximal entropy production",
{\sl Phys.~Rev.} {\bf A63} (2001) 022105, (quant-ph/0007111).


\bibitem{go-vl0006067}Gerald~A.~Goldin and Vladimir~Shtelen,
``On Galilean invariance and nonlinearity in electrodynamics and quantum mechanics",
quant-ph/0006067.

\bibitem{sc-mi0008108}A.~J.~Scott and G.~J.~Milburn,
``Quantum nonlinear dynamics of continuously measured systems",
{\sl Phys.~Rev.} {\bf A63} (2001) 042101, (quant-ph/0008108).

\bibitem{us-cz0011013}N.~V.~Ustinov, M.~Czachor, ``Darboux-integrable equations with non-Abelian nonlinearities", in {\sl Probing the structure of quantum mechanics: Nonlinearity,
nonlocality,  computation, axiomatics\/}, D.~Aerts, M.~Czachor, T.~Durt (Editors), 335 (World Scientific, Singapore, 2002), (nlin.SI./0011013).

\bibitem{br-hu0011125}D.~C.~Brody and L.~P.~Hughston,
``Stochastic Reduction in Nonlinear Quantum Mechanics",
{\sl Proc.~R.~Soc.~London} {\bf A458} (2002) 1117, (quant-ph/0011125).

\bibitem{miel0012041}Bogdan~Mielnik,
``Comments on: `Weinberg's Nonlinear Quantum Mechanics and Einstein-Podolsky-Rosen paradox', by Joseph Polchinski", quant-ph/0012041.


\bibitem{gheo0102110}S.~Gheorghiu-Svirschevski,
``Addendum to `Nonlinear quantum evolution with maximal entropy production'", {\sl Phys.~Rev.} \textbf{A63}, (2001) 054102, (quant-ph/0102110).

\bibitem{si-bu-gi0102125} Christoph~Simon, Vladimir~Buzek, and Nicolas Gisin,
``The no-signaling condition and quantum dynamics"'
{\sl Phys.~Rev.~Lett.} {\bf 87}, 170405 (2001), (quant-ph/0102125).

\bibitem{cz-do0106051}M.~Czachor and H.-D.~Doebner,
``Schroedinger-picture correlation functions for nonlinear evolutions",
quant-ph/0106051.

\bibitem{davi0106124}Mark~P.~Davidson,
``Comments on the nonlinear Schrodinger equation",
quant-ph/0106124.



\bibitem{cz-do0110008}Marek~Czachor and H.-D.~Doebner
``Correlation experiments in nonlinear quantum mechanics",
{\sl Phys.~Lett.} {\bf A301} (2002) 139, (quant-ph/0110008).

\bibitem{horo0111036}Pawel~Horodecki,
``From limits of quantum nonlinear operations to multicopy entanglement witnesses and state spectrum estimation",
quant-ph/0111036.

\bibitem{pard0111105}Miroslav~Pardy
``To the nonlinear quantum mechanics", quant-ph/0111105.

\bibitem{bere0112046} Gian-Paolo~Beretta ``Maximal-entropy-production-rate nonlinear quantum dynamics compatible with second law, reciprocity, fluctuation-dissipation, and time-energy uncertainty relations", quant-ph/0112046.

\bibitem{bona0201002}Pavel~Bona,
``Comment on `No-Signaling Condition and Quantum Dynamics'",
{\sl Phys. Rev. Letters} {\bf 90} (2003) 208901,
(quant-ph/0201002).



\bibitem{go-sh0201004}Gerald~A.~Goldin and Vladimir~M.~Shtelen,
``On gauge transformations of B\"acklund type and higher order nonlinear Schr\"odinger equations",
 {\sl  J.~Math.~Phys.\/} \textbf{43} (2002)  2180,
(quant-ph/0201004).

\bibitem{leif0201039}P.~Leifer,
``Super-Relativity and State-Dependent Gauge Fields",
gr-qc/0201039.

\bibitem{ca-ma-ro0202026}Carlos Castro, Jorge Mahecha, and Boris Rodriguez,
``Nonlinear QM as a fractal Brownian motion with complex diffusion constant", quant-ph/0202026.


\bibitem{ci-ga-ma0202076}R.~Cirelli, M.~Gatti, A.~Mani\'a, ``The Pure State Space of Quantum Mechanics as Hermitian Symmetric Space",  quant-ph/0202076.

\bibitem{mi-tz0202173}D.~Minic and C.~H.~Tze,
``Nambu Quantum Mechanics: A Nonlinear Generalization of Geometric Quantum Mechanics",
{\sl Phys.~Lett.} {\bf B536} (2002) 305, (hep-th/0202173).

\bibitem{gh-sv0203153}S.~Gheorghiu-Svirschevski,
``Quantum nonlocality and quantum dynamics",
quant-ph/0203153.

\bibitem{kent0204106}Adrian~Kent,
``Nonlinearity without Superluminality", quant-ph/0204106.


\bibitem{ci-cz-us0205061}J. L. Cieslinski, M. Czachor, N. V. Ustinov, ``Darboux covariant equations of von  Neumann type and their generalizations", {\sl J.~Math.~Phys.\/} {\bf 44} 1763 (2003), (nlin.SI/0205061).

\bibitem{bi-ni0205152}Jie~Liu, Biao~Wu, and  Qian~Niu,
``Adiabatic Theory of Nonlinear Evolution of Quantum States",
{\sl Phys.~Rev.~Lett.} {\bf 90} (2003) 170404, (quant-ph/0205152).

\bibitem{ae-va0205162}Diederik Aerts, Frank Valckenborgh,  ``The Linearity of Quantum Mechanics at Stake: The Description of Separated Quantum Entities",
 In {\sl Probing the Structure of Quantum Mechanics: Nonlinearity, Nonlocality, Computation and  Axiomatics\/}, eds.
D.~Aerts, M.~Czachor and T.~Durt, World Scientific, Singapore (2002)  p. 20, (quant-ph/0205162).

\bibitem{ae-va}Diederik~Aerts, Frank~Valckenborgh, ``Linearity and Compound Physical Systems: The Case of Two Separated Spin 1/2 Entities",  In {\sl Probing the Structure of Quantum Mechanics: Nonlinearity, Nonlocality, Computation and  Axiomatics\/}, eds.
D.~Aerts, M.~Czachor and T.~Durt, World Scientific, Singapore (2002) p. 47, (quant-ph/0205166).


\bibitem{fu-ts-bo0207024}Wilhelm~I.~Fushchych, Ivan~M.~Tsyfra, Vyacheslav~M.~Boyko,  ``Nonlinear representations for Poincare and Galilei algebras and nonlinear equations for electromagnetic fields", {\sl J.~Nonlin.~Math.~Phys.\/} {\bf  2} 210 (1994), (math-ph/0207024).

\bibitem{gheo0207042}S.~Gheorghiu-Svirschevski,
``A General Framework for Nonlinear Quantum Dynamics", quant-ph/0207042.

\bibitem{do-do-tw0207077}V.~K.~Dobrev, H.-D.~Doebner, R.~Twarock,  ``Quantum Mechanics with Difference Operators", {\sl Rep.~Math.~Phys.\/} {\bf 50} 409 (2002), (quant-ph/0207077).


\bibitem{fu-bo020816}Wilhelm~I.~Fushchych, Vyacheslav~M.~Boyko,  ``Continuity Equation in Nonlinear Quantum Mechanics and the Galilei Relativity Principle",
{\sl J.~Nonlin.~Math.~Phys.\/} {\bf 4} 124 (1997), (math-ph/0208016).


\bibitem{svet0208049}George~Svetlichny,
  ``Critique of `No-Signaling Condition and Quantum Dynamics'",
quant-ph/0208049.

\bibitem{ka-sc0209075}G.~Kaniadakis and A.~M.~Scarfone,
``Nonlinear gauge transformation for a class of Schroedinger equations containing complex nonlinearities",
{\sl Rep.~Math.~Phys.} {\bf 46} 113 (2000), (quant-ph/0209075).

\bibitem{ka-sc0209130}G.~Kaniadakis and A.~M.~Scarfone,
``Nonlinear Transformation for a Class of Gauged Schroedinger Equations with Complex Nonlinearities",
{\sl Rep.~Math.~Phys.} {\bf 48} 115 (2001), (quant-ph/0209130).

\bibitem{od-ra0210207}Anatol~Odzijewicz, Tudor~S.~Ratiu, ``Banach Lie-Poisson spaces and reduction",  {\sl Commun.~Math.~Phys.\/} {\bf 243} 1 (2003), (math.SG/0210207).

\bibitem{ae-cz-ga-etal0211105}D.~Aerts, M.~Czachor, L.~Gabora, M.~Kuna, A.~Posiewnik,
J.~Pykacz, M.~Syty,
``Quantum morphogenesis:  A variation on Thom's ctastrophe theory", {\sl Phys. Rev.\/} {\bf E67} 051926 (2003);  {\sl Virtual Journal of Biological Physics\/}, June 2003, (quant-ph/0211105).

\bibitem{zi-st0211149} Mario Ziman and Peter Stelmachovic ``Quantum theory: kinematics, linearity and no-signaling condition", quant-ph/0211149.

\bibitem{ni-po0301009}Anatoly G.~Nikitin and Roman O.~Popovych,
``Group classification of nonlinear Schr\"odinger equations",
{\sl Ukr.~Math.~J.} {\bf 53} (2001) 1255, (math-ph/0301009).

\bibitem{wang0304101}Charles Wang,
``A nonlinear quantum model of the Friedmann universe",
{\sl Class.~Quant.~Grav.} {\bf 20} (2003) 3151, (gr-qc/0304101).

\bibitem{do-zh0304167}H.-D.~Doebner and R.~Zhdanov,
``Nonlinear Dirac equations and nonlinear gauge transformations",
quant-ph/0304167.

\bibitem{svet0305100}George~Svetlichny,
``Non-linear quantum mechanics and high energy cosmic rays",
{\sl Found.~Phys.~Lett.} {\bf 17} (2004) 197, (hep-th/0305100).

\bibitem{sing0306110}T.~P.~Singh,
``Quantum mechanics without spacetime III: a proposal for a non-linear Schrodinger equation", gr-qc/0306110.


\bibitem{kalb0307018}G.~K\"albermann,
``Ehrenfest theorem, Galilean invariance and nonlinear Schr\"odinger equations",
quant-ph/0307018.

\bibitem{yala0307116}M.~Cemal~Yalabik,
``Nonlinear Schrodinger Equation for Quantum Computation",
quant-ph/0307116.

\bibitem{svet0308001}George~Svetlichny,
``On linearity of separating multi-particle differential Schr\"odinger operators for identical particles",
{\sl Journal of Mathematical Physics} {\bf 45} 959 (2004), (quant-ph/0308001).



\bibitem{wang0308026}Charles Wang,
``Nonlinear parametric quantization of gravity and cosmological models",
gr-qc/0308026.

\bibitem{ae-cz-sy0310009}D.~Aerts, M.~Czachor, M.~Syty, ``Quantum circa rhythms", physics/0310009.


\bibitem{wa-ke-ir0310129}Charles~H-T Wang, Smaragda~Kessari, and Edward~R.~Irvine,
``An action principle for the quantization of parametric theories and nonlinear quantum cosmology",
{\sl Classical and Quantum Gravity} {\bf 21} (2004) 2319, (gr-qc/0310129).

\bibitem{gh-sv0311042}S.~Gheorghiu-Svirschevski,
``Inferring linear quantum dynamics without no-signaling",
quant-ph/0311042.

\bibitem{go-pr0311269}Daniel Gottesman and John Preskill, ``Comment on `The black hole final state'\,", {\sl JHEP} {\bf 0403} 026 (2004) (hep-th/0311269).

\bibitem{yu-ho0312160}Ulvi~Yurtsever and George~Hockney, ``Causality, Entanglement, and Quantum Evolution Beyond Cauchy Horizons", quant-ph/0312160.

\bibitem{aaron0401062}Scott~Aaronson,
``Is Quantum Mechanics An Island In Theoryspace?",
quant-ph/0401062.

\bibitem{gesz0401086} Tamas~Geszti, ``Gravitational self-localization in quantum measurement", {\sl Phys.~Rev.\/} {\bf A69} 0321101 (2004), (quant-ph/0401086).

\bibitem{parw0401190}Rajesh~R.~Parwani,  ``An Information-Theoretic Link Between Spacetime Symmetries and Quantum Linearity", {\sl Annals Phys.\/} \textbf{315} 419 (2005), (hep-th/0401190).

\bibitem{yu-ho0402060} Ulvi~Yurtsever and George~Hockney, ``Signaling and the Black Hole Final State", hep-th/0402060.

\bibitem{bere0402180}Gian~Paolo~Beretta,
``Nonlinear extensions of Schr\"odinger-von Neumann quantum dynamics: a list of conditions  for compatibility with thermodynamics", {\sl Modern Physics Letters A\/} \textbf{20} 977--984 (2005), (quant-ph/0402180).

\bibitem{wang0406079}Charles~ H.-T.~Wang,
``Nonlinear quantum gravity on the constant mean curvature foliation",
gr-qc/0406079.

\bibitem{parw0408185}Rajesh~R.~Parwani,  ``Information Measures for Inferring Quantum Mechanics and its Deformations",  quant-ph/0408185.

\end{thebibliography}
\end{document}